\pdfoutput=1
\documentclass[10pt,conference]{IEEEtran}
\IEEEoverridecommandlockouts
\usepackage{graphicx} 
\usepackage{lipsum} 
\usepackage{amsmath}
\usepackage[dvipsnames]{xcolor}

\usepackage{tikz}
\usepackage{yquant}
\usetikzlibrary{quantikz}
\usepackage{braket}
\usepackage{subcaption}
\usepackage{csquotes}
\usepackage{hyperref}
\usepackage{cleveref}
\hypersetup{%
linktocpage=true, 
colorlinks=false,
pdfborder={0 0 0}, 
breaklinks=true, pdfpagemode=UseNone, pageanchor=true, pdfpagemode=UseOutlines,%
plainpages=false, bookmarksnumbered, bookmarksopen=true, bookmarksopenlevel=1,%
hypertexnames=true, pdfhighlight=/O,
pdftitle={CutReg: A loss regularizer for enhancing the scalability of QML via adaptive circuit cutting},%
pdfauthor={Maniraman Periyasam},%
pdfsubject={QML},%
pdfkeywords={},%
pdfcreator={pdfLaTeX},%
pdfproducer={LaTeX with hyperref}%
}

\begin{document}

\title{CutReg: A loss regularizer for enhancing the scalability of QML via adaptive circuit cutting}
\author{
    \IEEEauthorblockN{Maniraman Periyasamy\IEEEauthorrefmark{1}\IEEEauthorrefmark{2}, Christian Ufrecht\IEEEauthorrefmark{1} Daniel D.\ Scherer\IEEEauthorrefmark{1}, Wolfgang Mauerer\IEEEauthorrefmark{2}\IEEEauthorrefmark{3}
    }
    \IEEEauthorblockA{
        \IEEEauthorrefmark{1}Fraunhofer IIS, Fraunhofer Institute for Integrated Circuits IIS, Nürnberg, Germany\\
        \IEEEauthorrefmark{2}Technical University of Applied Sciences Regensburg, Regensburg, Germany\\
        \IEEEauthorrefmark{3}Siemens AG, Corporate Research, Munich, Germany
    }
    \thanks{email address for correspondence: daniel.scherer2@iis.fraunhofer.de}
}

\maketitle

\begin{abstract}
Whether QML can offer a transformative advantage remains an open question. The severe constraints of NISQ hardware, particularly in circuit depth and connectivity, hinder both the validation of quantum advantage and the empirical investigation of major obstacles like barren plateaus. Circuit cutting techniques have emerged as a strategy to execute larger quantum circuits on smaller, less connected hardware by dividing them into subcircuits. However, this partitioning increases the number of samples needed to estimate the expectation value accurately through classical post-processing compared to estimating it directly from the full circuit. This work introduces a novel regularization term into the QML optimization process, directly penalizing the overhead associated with sampling. We demonstrate that this approach enables the optimizer to balance the advantages of gate cutting against the optimization of the typical ML cost function. Specifically, it navigates the trade-off between 
minimizing the cutting overhead and maintaining the overall accuracy of the QML model, paving the way to study larger complex problems in pursuit of quantum advantage. 
\end{abstract}

\section{Introduction}

\IEEEPARstart{Q} Machine Learning (QML) seeks to leverage the principles of quantum mechanics to enhance machine learning algorithms, potentially offering speedups or improved model performance for specific tasks. However, the capabilities of current Noisy Intermediate-Scale Quantum (NISQ) devices are constrained by limited qubit counts, coherence times, and gate fidelities. These limitations hinder the direct implementation of complex quantum circuits often required for QML models. To address these hardware constraints, \enquote{gate cutting} (also known as circuit knitting or circuit partitioning) techniques have gained significant attention \cite{Hofmann_2009, Peng_2020, Mitarai2021}. Gate cutting is a method to partition a circuit into smaller subcircuits by replacing the multi-qubit rotation gates between two qubits with different single-qubit gates. The measurement results executing subcircuits are then classically combined, but this process introduces a significant sampling overhead \cite{Peng_2020}. This overhead is highly sensitive to the parameters of the cut gates, often called cutting angles. Angles that minimize the overhead often correspond to reduced entanglement. This creates a potential trade-off between the expressivity of the ansatz and the computational cost. We propose a novel regularization term within the QML model's loss function to penalize the estimated sampling overhead. This encourages the optimizer to find solutions that balance the primary learning task with a manageable sampling budget, automating the navigation of the complex interplay between circuit partitioning and classical post-processing costs.

\section{Literature Review}

The concept of gate cutting, or circuit knitting, originates from the broader field of quantum circuit simulation and compilation. Foundational work by Hofmann \cite{Hofmann_2009} established a formal framework for partitioning circuits. Their work was crucial in highlighting the primary challenge: an exponential scaling of the classical post-processing cost and sampling overhead with the number of cuts. Subsequent research focused on understanding and mitigating this overhead. Building on this, Peng et al. \cite{Peng_2020} provided a rigorous and detailed analysis of the sampling overhead, formally quantifying the overhead factor and introducing improved circuit-cutting methods by minimizing associated classical computing costs. Mitarai et al., \cite{Mitarai2021} explored techniques for gate cutting, and importantly, formulated the sampling cost required for gate cutting. 
In the context of QML, the application of these techniques is relatively nascent but highly relevant. Many leading QML algorithms, such as Variational Quantum Algorithms (VQAs)\cite{Periyasamy2022}, rely on parameterized quantum circuits, making the interplay between trainable circuit parameters and gate-cutting parameters a crucial area of study. However, existing approaches typically treat gate cutting as a static pre-processing step, where cuts are determined heuristically based on hardware topology before training begins \cite{Marshall_2023}. Regularization is a standard and powerful technique in classical machine learning used to prevent overfitting and incorporate prior knowledge or constraints into an optimization problem. Gentinetta et al. \cite{Gentinetta_2024} developed a method to constrain the sampling overhead for simulating a dynamic quantum system for chemical simulations. However, to the best of our knowledge, the direct application of a regularizer to penalize the sampling overhead induced by gate cutting during the generic QML optimization process has not been previously investigated. We leverage the detailed understanding of overhead mechanics from \cite{Peng_2020} and the insights into parameter-dependent costs from \cite{Mitarai2021} to create a novel, adaptive component within the QML training pipeline. This shifts gate cutting from a static step to a dynamic, trainable process that allows the optimizer to find a practical balance between model performance and the quantum resources required for its execution.

\section{Proposed Method} 
\label{sec:proposed_method}

 We introduce a novel strategy to mitigate the substantial sampling overhead associated with gate cutting in QML applications. Our approach integrates a regularization term directly into the QML model's optimization objective, thereby enforcing the learning algorithm to find a balance between achieving the primary learning task and minimizing the sampling overhead arising from circuit cutting.
The core principle of gate cutting involves decomposing a target quantum gate $U$, typically a two-qubit gate that exceeds hardware connectivity or depth constraints, into a set of classically correlated, simpler operations. Consider a generic two-qubit unitary gate $U$ that we want to decompose. Techniques such as gate decompositions effectively replace $U$ with a sum of operations that can be executed on smaller subcircuits \cite{Peng_2020, Ufrecht_2023, Ufrecht_2024}. 
Gate cutting is the task of finding a decomposition for the unitary channel $\mathcal{U}$ such that $\mathcal{U}=\sum_k c_k \mathcal{F}_k$. Here, $\mathcal{F}_k$ are local channels operating on a single or sub-group of qubits, and $c_k$ are the coefficients. Some of the coefficients, $c_k$ necessarily have to be negative to allow the simulation of non-local gates by local operations and give rise to a quasi-probability distribution $p_k = |c_k| / \sum_j |c_j|$. The circuit execution then involves random execution of the subcircuits containing the local operator $\mathcal{F}_k$ with probability $p_k$ and weighting the outcome by $\text{sgn}(c_k)$. Classical post-processing of the results from the sub-circuits then obtains the expectation value of an observable on the whole circuit. Quasiprobability simiulation increases the variance of the expectation value measurement of the observable, necessitating a larger random execution of the subcircuits and the averaging of the results, and the post-processing function induces an additive error. This error necessitates a larger number of shots to achieve a level of accuracy comparable to that obtained from executing the whole uncut circuit. This requirement of a larger number of shots due to the cutting procedure is referred to as sampling overhead $s$ and is quantified by $\mathcal{O}(\kappa^2)$ where $\kappa = \Sigma_k|c_k|$.

Here, if we consider a two-qubit non-local gate $U(\alpha)$ parameterized by an angle $\alpha$, for example, $Rzz(\alpha)$, the cutting procedure transforms it into an angle-dependent quasi-probabilistic mixture $\mathcal{U}=\sum_k c_k(\alpha) \mathcal{F}_k$ and sampling overhead $s(\alpha)$. The $p_k(\alpha)$ dependent on $c_k(\alpha)$ is the angle-dependent quasi-probabilities. For cutting the $Rzz(\alpha)$ gate, the angle-dependent sampling overhead is known and given by the  \cref{eq:rzz_sampling_overhead} below \cite{Piveteau2022_circuitcut}. The $s(\alpha)$ shows minimal overhead when $\alpha$ is near $0$ or $\pi$, but substantial overhead around $\alpha = \frac{\pi}{2}$, which corresponds to maximal entanglement generation capability by the original gate. When the circuit involves $L$ number of $Rzz$ gates to be cut, the total sampling overhead is given by \cref{eq:mult_rzz_sampling_overhead} below where $\hat{\alpha} = (\alpha_1, \dots, \alpha_L)$ is the vector of cutting angles.
\begin{equation}
    \label{eq:rzz_sampling_overhead}
    s(\alpha) = (1+2|\text{sin}(\alpha)|)^2
\end{equation}
\begin{equation}
    \label{eq:mult_rzz_sampling_overhead}
    s(\hat{\alpha}) = \prod_{l=1}^{L}(1+2|\text{sin}(\alpha_l)|)^2
\end{equation}

In a standard VQA or QML framework, a parameterized quantum circuit $U_C(\hat{\theta})$ is optimized to minimize a loss function $L_{QML}(\hat{\theta})$, which typically depends on the expectation value of an observable $\boldsymbol{O}$: $\hat{\theta}^* = \arg\min_{\hat{\theta}} L_{QML}(\langle \boldsymbol{O} \rangle_{\hat{\theta}})$. If gate cutting is applied, some parameters within $\hat{\theta}$ may correspond to the cutting angles $\alpha_l$ for $L$ distinct cuts within $U_C(\hat{\theta})$. Each cut $l$ introduces an overhead $s_l(\alpha_l)$.
Our main proposal is to adjust this loss function to incorporate a penalty for the cumulative sampling overhead. The modified regularized objective function $L_{\text{reg}}(\hat{\theta})$ is:
\begin{equation} 
L_{\text{reg}}(\hat{\theta}) = L_{QML}(\langle \boldsymbol{O} \rangle_{\hat{\theta}}) + \lambda \cdot R_{\text{overhead}}(\hat{\alpha})
\end{equation}
Here, the cutting angles $\hat{\alpha}$ is part of $\hat{\theta}$, $\lambda \ge 0$ is a hyperparameter controlling the regularization strength, and $R_{\text{overhead}}(\hat{\alpha})$ is the regularization term, which is a function of the total sampling cost. The term $R_{\text{overhead}}(\hat{\alpha})$ should be a non-negative, differentiable function that increases monotonically with the sampling overhead from each cut. 
For instance, considering $L$ independent cuts, a practical regularizer can be formulated as the logarithm of the product of the individual overheads given by \cref{eq:reg_overhead}.
\begin{equation}
\label{eq:reg_overhead}
    R_{\text{overhead}}(\hat{\alpha}) = \log \left( \prod_{l=1}^{L}(1+2|\text{sin}(\alpha_l)|)^2 \right)
\end{equation}
This form provides a direct penalty for increased sampling overhead. Other forms, such as those based on the sum of $s(\alpha_l)$ values (e.g., $ R_{\text{overhead}}(\hat{\alpha}) = \sum_{l=1}^{L} (s(\alpha_l) - 1) $, where $s(\alpha_l)=1$ implies no overhead penalty from cut $l$), could also be considered. In this paper, we adopt the formulation shown in \cref{eq:reg_overhead} due to its simplicity and scaling behavior for larger sampling overheads.

The optimization of $L_{\text{reg}}(\hat{\theta})$ aims to find the circuit parameters $\hat{\theta}^*$, which contain $\hat{\alpha}^*$, that not only perform well on the QML task (minimizing $L_{QML}$) but also maintain low values for the total sampling overhead. The hyperparameter $\lambda$ controls this balance: a smaller $\lambda$ prioritizes QML performance, potentially tolerating higher sampling costs, while a larger $\lambda$ imposes a strong penalty on overhead, possibly guiding cutting angles towards $0$ or $\pi$ even if it slightly compromises the QML objective by reducing entanglement. This method allows the VQA to autonomously adapt the cutting strategy (via the parameters $\alpha_l$) during the training process, making the choice of cutting parameters an integral part of the learning problem. 

\section{Experimentation and Results}
\label{sec:experimentation_results}

To evaluate the efficacy of the proposed regularization method for managing sampling overhead in gate-cut QML circuits, we conducted a series of numerical experiments focusing on a regression task. We investigated the interplay between model accuracy, sampling overhead, and entangling capability across varying qubit numbers.

\subsection{Experimental Setup}
\subsubsection{QML Regression Task and Dataset}
The QML model was trained to learn a synthetic regression function generated using the $make\_regression$ function from scikit-learn. A training dataset consisting of $N_{train} = 100$ samples, a separate validation set of $N_{val} = 50$ samples, and a test set of $N_{test} = 50$ samples was created for training and evaluation. The dataset was designed so that the number of features is twice the number of qubits in the QML model. A new dataset was generated for each QML model with a different number of qubits in the VQCs.

\subsubsection{Quantum Circuit Ansatz and Gate Cutting}
\begin{figure}[htbp!]
    \centering
    \resizebox{0.85\linewidth}{!}{
    \begin{tikzpicture}
        \input{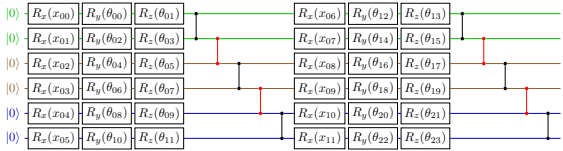}
    \end{tikzpicture}
    
    }
    \caption{The 6-qubit VQC ansatz used as the function approximator of the QML model. The qubits are color-coded based on the partitions, and the $CZ$ gates to be cut are marked in red.}
    \label{fig:vqc}
\end{figure}

\begin{figure}[htbp!]
    \centering
    \resizebox{0.5\linewidth}{!}{
    \begin{tikzpicture}
        \input{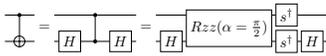}
    \end{tikzpicture}
    
    }
    \caption{Decomposition of $CX$ gate into equivalent $CZ$ and $Rzz$ representation}
    \label{fig:CR-Rzz}
\end{figure}

We employed a hardware-efficient ansatz composed of layers of single-qubit rotations ($R_y(\theta), R_z(\phi)$) and entangling blocks of CZ gates arranged in a nearest neighbour topology. Each feature of input $x$ was encoded into the VQC using $R_x$ rotations and incremental data-uploading \cite{Periyasamy2022}. The total number of layers was scaled as required based on number of qubits in the VQC and the number of input features. All the VQCs evaluated in this work were divided into three sub-circuits using gate cutting technique as shown in \cref{fig:vqc}. To aid the regularization and reduction in sampling overhead as explained in the section \ref{sec:proposed_method}, the $CZ$ gates that were cut are replaced by trainable $Rzz$ gates as shown in the \cref{fig:CR-Rzz}. The cutting parameters (angles $\alpha_l$ used in the decomposition of the cut gates, as described in Section \ref{sec:proposed_method}) were treated as trainable variables, initialized to $\frac{\pi}{2}$ to represent an initially unbiased cut.

\subsubsection{Simulation and Optimization}
All experiments were conducted using the Qiskit quantum simulator and Qiskit add-on: circuit cutting \cite{qiskit-addon-cutting}. The parameters of the QML model, denoted by $\hat{\theta}$, which include both the single qubit gate parameters and the cutting angles $\hat{\alpha}$, were optimized using gradients estimated via the guided-SPSA technique introduced in Ref.~\cite{periyasamy_2024_guidedspsa} and the Amsgrad optimizer \cite{Wiedmann2023}. This optimization was performed with a learning rate of $\eta = 0.01$ over a total of $N_{epochs} = 100$ epochs and batch size of 32. The loss function employed for the regression task was the Mean Squared Error (MSE), augmented with the proposed sampling overhead regularizer expressed as: 
\begin{equation}
    L_{\text{reg}} = \text{MSE}(\hat{\theta}, \hat{\alpha}) + \log \left( \prod_{l=1}^{L}(1+2|\text{sin}(\alpha_l)|)^2 \right)
\end{equation}
The initial value of $\lambda$ was set to 0.01 and stepped down to 0.0001 after 10 epochs to ensure that the regularizer's penalty does not overpower the actual training loss. This approach allows the trained model to end up with larger sampling overhead if needed for optimal solution.

\begin{figure*}[tb]
    \centering

    \begin{subfigure}{0.3\textwidth}
        \centering
        \vspace*{-1em}\input{plots/18_test.tex}\vspace*{-2em}
        \caption{18-Qubit VQC}
        \label{fig:18_vqc_test}
    \end{subfigure}%
    \hfill
    \begin{subfigure}{0.3\textwidth}
        \centering
        \vspace*{-1em}\input{plots/sample_overhead_18.tex}\vspace*{-2em}
        \caption{18-Qubit VQC}
        \label{fig:18_vqc_sample}
    \end{subfigure}%
    \hfill
    \begin{subfigure}{0.3\textwidth}
        \centering
        \vspace*{-1em}\input{plots/24_test.tex}\vspace*{-2em}
        \caption{24-Qubit VQC}
        \label{fig:24_vqc_test}
    \end{subfigure}%
    \hfill
\end{figure*}

\subsubsection{Metrics}
The primary performance metric for the regression task was the mean squared error (MSE) on the test set. The total sampling overhead $S_{total} = \prod_l s(\alpha_l)$ was calculated at the start of every epoch during the training. To quantify entanglement, we computed the Meyer-Wallach multipartite entanglement measure $Q$ \cite{brennen2003observablemeasureentanglementpure} for the state prepared by the quantum circuit before measurement. The $Q$ measure is defined as $Q = 2(1-\frac{1}{N_q} \sum_{k=1}^{N_q} (1 - \text{Tr}(\rho_k^2))$, where $\rho_k$ is the reduced density matrix of qubit $k$. The value of $Q$ ranges from $0$ (fully separable state) to $1$ (certain maximally entangled states). The entanglement measure was also tracked during the course of training.

\subsection{Results and Analysis}
We conducted experiments for quantum circuits with $N_q = 18$, $24$, $30$ and $50$ qubits.

\subsubsection{Regression Performance}

The introduction of the regularizer generally allowed the QML model to maintain good regression performance while actively managing the cutting angles. The \cref{fig:18_vqc_test} presents the average test results obtained from ten training runs of the 18-qubit VQC, where the angles of the $Rzz$ gates designated for cutting are initialized to $\frac{\pi}{2}$, corresponding to maximum entanglement generation, and to $0.1$, corresponding to partial entanglement generation capability of each individual gate. The \cref{fig:18_vqc_sample} illustrates the corresponding changes in sampling overhead for both models throughout the training process. The maximally entangled model began with a sampling overhead of 6561 to cut the four $Rzz$ gates, while the partially entangled model had a sampling overhead of 4.29 for the same number of cut gates. Nevertheless, by the end of the training, both models resulted in a similar sampling overhead of about 10, as shown in \cref{fig:18_vqc_sample}, and achieved comparable results, as demonstrated in \cref{fig:18_vqc_test}. 
By starting with gates with reduced entangling capability for the cutting procedure, we can gain a computational advantage that allows us to simulate results with higher accuracy due to the small sampling overhead.  The \cref{fig:24_vqc_test} depicts the test performance of the 24 qubits VQC model. The convergence here is slow compared to the 18 qubit circuits and longer training time is needed. The cutting procedure made training 30 and 50-qubit VQC models possible on a single CPU with 50GB RAM. However, the training procedure resulted in gradients in the order of $10^-7$, indicating that the models have hit the barren plateau regime. 

\begin{figure}[htbp!]
\label{fig:entanglement}
    \centering
    \vspace*{-2em}\input{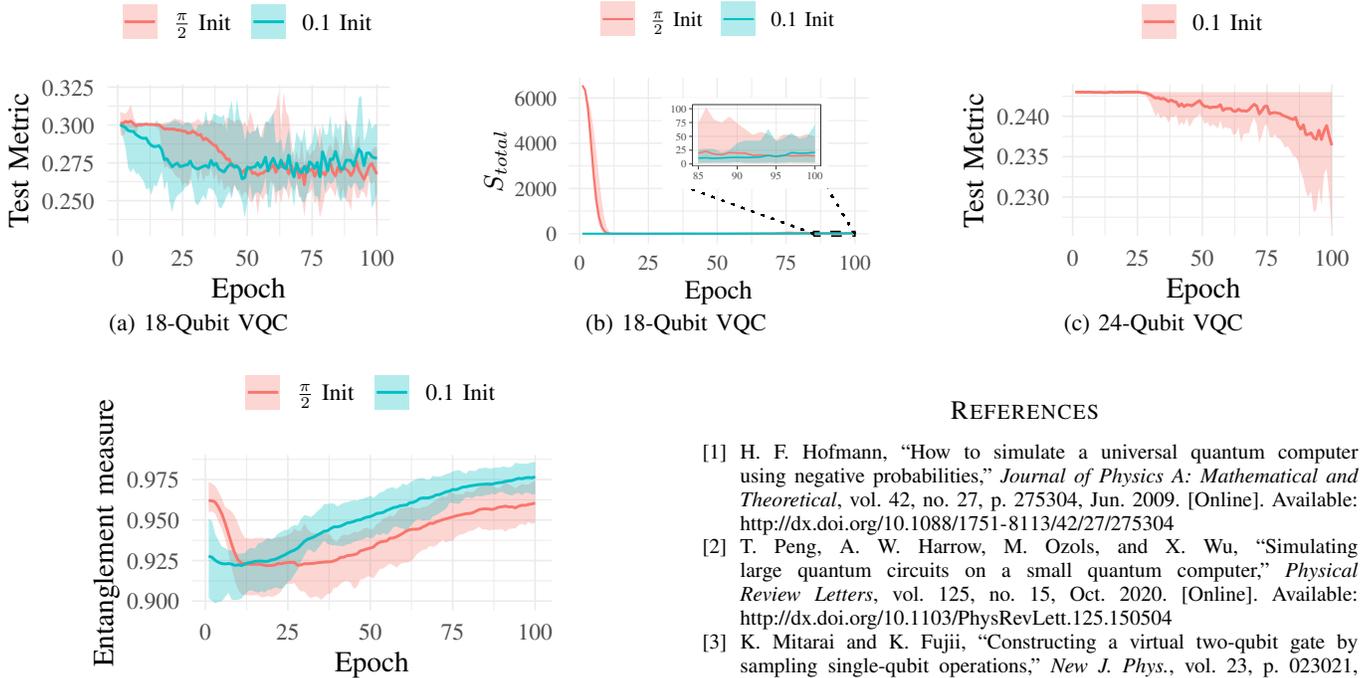}\vspace*{-1em}
    \caption{Overall entanglement created by the VQC over training epochs}
\end{figure}

\subsubsection{Meyer-Wallach Entanglement Measure}

The \cref{fig:entanglement} illustrates that for an 18-qubit VQC, both maximally and partially entangled models initially exhibit a decrease in overall entanglement during the early stages of training. However, as training progresses, the entanglement surpasses its initial value. This suggests that the VQC's parameters are being optimized to enhance the total entanglement despite the presence of gates that generate partial entanglement between different partitions of the system.

\section{Conclusion}
\label{sec:conclusion}

In this work, we addressed the significant challenge of sampling overhead in circuit cutting by introducing a new regularization term into the quantum machine learning loss function. This method encourages the optimizer to achieve a careful balance between model accuracy and the traditional post-processing budget. Our numerical experiments on a regression task, conducted with Variational Quantum Circuits of up to 50 qubits, validate the efficacy of our approach, demonstrating for an 18-qubit regression task that models initialized with gates that generate maximal and partial entanglement converged to a similar state of accuracy and sampling overhead. This key finding indicates that maximal entanglement at the circuit cuts is not necessary for the optimal performance of the gate cutting-based QML model and that our method can automatically achieve this balance. While our method enabled the training of larger 30 and 50-qubit models, the emergence of barren plateaus highlights that other scalability challenges persist. The proposed technique significantly enhances the practicality of QML on near-term hardware by automating the trade-off between circuit size and sampling cost. Specifically, the improved computational efficiency allows for the effective study of large-scale phenomena, such as barren plateaus, without excessive computational demands. Integrating gate cutting into the training pipeline enhances scalability for QML applications and future research that combines this overhead-aware training with barren plateau mitigation techniques.

\bibliographystyle{IEEEtran}
\bibliography{references}

\end{document}